\documentclass{article}
\usepackage{float}
\usepackage{amsmath}
\usepackage{amssymb}
\usepackage{graphicx}
\usepackage{esint}

\usepackage{algorithm}
\usepackage[noend]{algpseudocode}
\usepackage{hyperref}
\hypersetup{
    colorlinks=false,
}


\date{}

\begin{document}
\title{Skeleton based simulation of turbidite channels system}


\author{Viviana Lorena Vargas, Sinesio Pesco\\
  \small Pontificia Universidade Catolica, Rio de Janeiro\\
  \small vivagra@puc-rio.br, sinesio@puc-rio.br\\
}

\maketitle


\begin{abstract}
A new approach to model turbidite channels using training images is presented, it is called \textbf{skeleton based simulation}. This is an object based model that uses some elements of multipoint geostatistics. The main idea is to simplify the representation of the training image by a one-dimensional object called training skeleton. From this new object, information about the direction and length of the channels is extracted and it is used to simulate others skeletons. These new skeletons are used to create a 3D model of channels inside a turbidite lobe.  

\end{abstract}

\section{Introduction} 

Geoscience is the study of the planet Earth and its different natural geologic systems. Frequently structures in the earth present a heterogeneous nature making the stationary modeling inappropriate. The study of oil reservoirs is part of the geoscience studies, and an important issue is the modeling of certain type of reservoirs whose image representation cannot be assumed as the realization of a stationary point process, however it has a well-defined geometric structure, like fan deltas and turbidites deposits.

\begin{figure}[h!]
	\centering 
    \includegraphics[width=9.5cm]{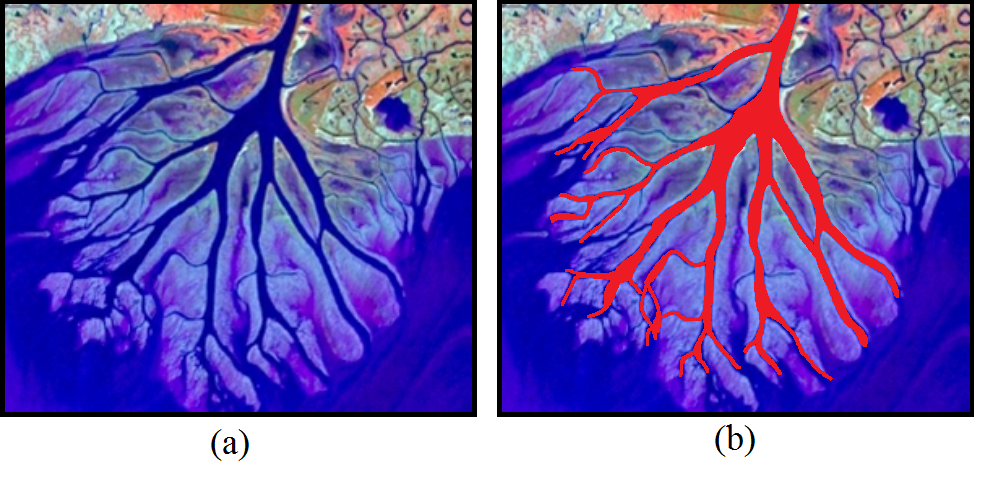}
    \caption{\small (a) The Wax Lake Delta. (b) The channel system highlighted in red is the training image.} \label{Wax}
\end{figure}

The purpose of this work is the modeling of geologic structures, which spatial continuity can be assumed to be reflected by training images like the red channel system in the figure \ref{Wax}(b); these images present a particular geo\-me\-try that we called \textbf{tree-like}. We proposed an object-based method called SKESIM. It born from the idea of interpreting tree-like images as the thickening of a graph.

Object-based methods describe the geological structure modeled by parametric geometries and it is represented by a co\-llec\-tion of well-defined geometric objects. Such geometric objects are defined by parameters deduced from the available information. Then, the simulation proceeds sequentially creating the objects and placing them in the simulation grid until a criteria is attained \cite{pyr}. 

SKESIM builds a 3D model of the channel system using information extracted from a tree-like image. This image is used as training image, that means as an example of the spatial continuity that is believed to be presented in a natural phenomena. It is a conceptual image that is assumed to contain all possible structures of interest believed to appear in the geological body \cite{MC}. The main idea is to represent the training image as a uni-dimensional object called \textit{skeleton}. This object is simple and contains the most basic information of the image. It is used to define probability distributions from which the parameter values of the new skeletons are sampled.

Specifically, we are interested in the modeling of turbidite channel systems. Turbidite deposits are generated by turbidity currents and related gravity flows \cite{PFC}. Turbidite reservoirs are distinguished by a complex structure of sand bodies arranged in channels and lobes \cite{zhang}. This kind of reservoirs still represent an important source of oil exploration in Brazil. The cost of drilling a single well can easily exceed 100 million dollars and the success rates are around 15 to 30 percent, then the risk involved in the exploitation must be determined. It is necessary an adequate representation of these reservoirs. These kind of structures have been studied in works like \cite{sullivan}, \cite{marques}, \cite{beau}. The main objective of this work is to present a methodology for modeling the depositional architecture of turbidite channels.

\begin{figure}[h!]
\centering
\includegraphics[width=8.5cm]{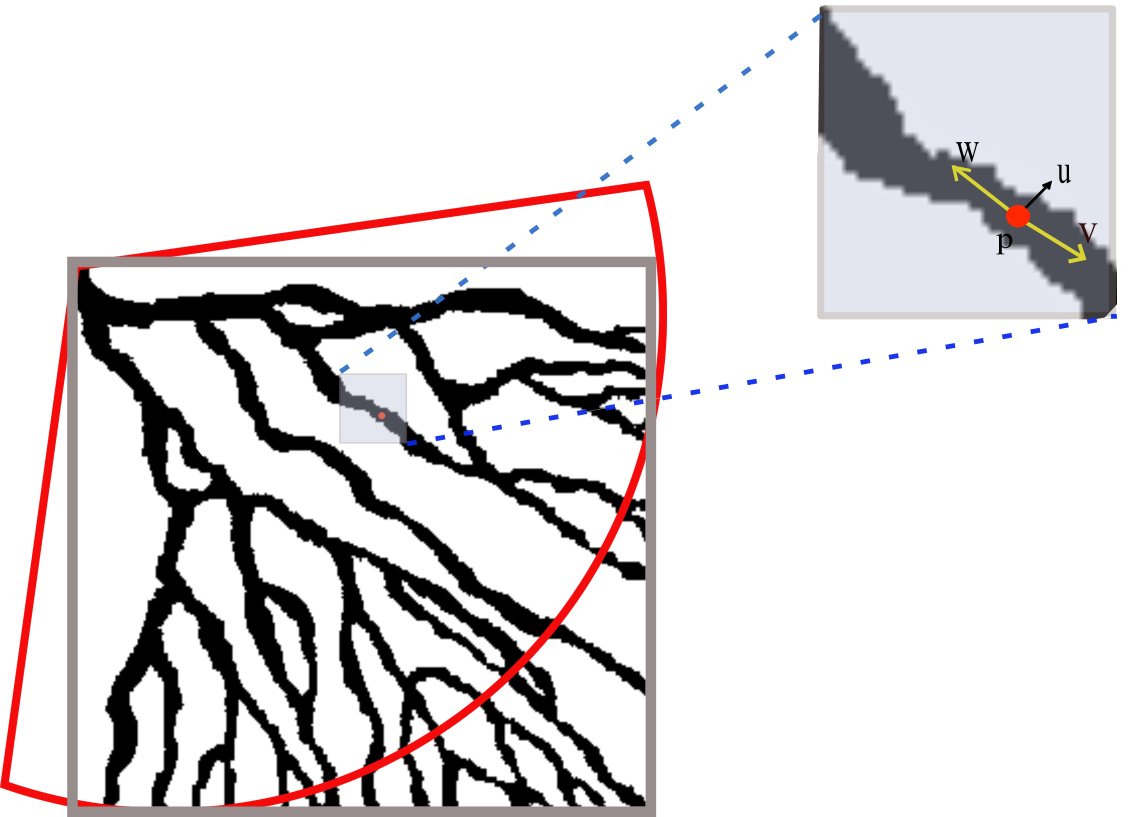}
\caption{Three directions are perceived, $ v $ and $ w $ in the channel development direction and $u$ in the width direction of the channel. Visual definition of the directional interval in a tree-like image.} \label{two_dir}
\end{figure}

The turbidite channel system that will be modeled is represented by binary 2D training images. Because of the particular structure of the modeled object, we restrict to images with a special geometry, that we called tree-like.  We have not formal definition, but tree-like images should have the following features: (1) At each black point, one direction in which the channel appears to be developing can be perceived. (2) The tree-like image has a directional interval, defined as an interval containing the directions that are presented in the image. It can be established visually, for example in figure \ref{two_dir} the directions in the tree-like are inside the red cone.

For the 3D modeling, the turbidite channel system is assumed to be located in the half-space $\{z\leq 0\}$ and the training image is thought as its projection to the plane $xy$. The channel system is interpreted as a thickening and deepening of the skeleton defined from the training image. It is created inside a turbidite lobe. 

\section{Skeleton based simulation}

The channel system is approximated by a one dimensional structure, that will be called \textit{skeleton}. It is a graph in the plane formed by edges that are straight lines. The nodes represent the channel bifurcations and all skeleton has a special node corresponding to the root of the graph. This skeleton reflects the global behavior of the channel system.

\begin{figure}[h!]
	  	\centering
		\includegraphics[width=10.5cm]{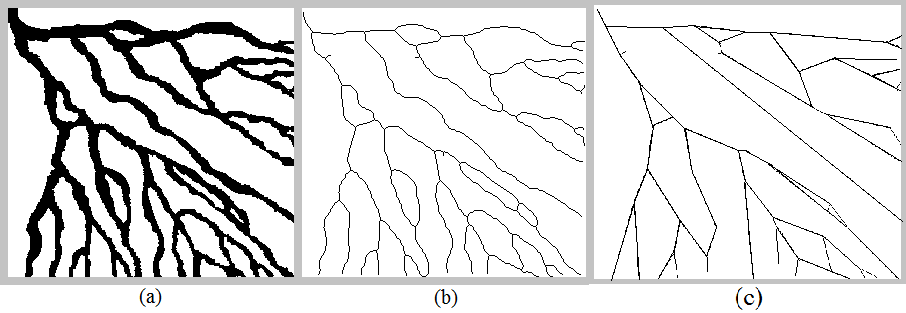}
		\caption{(a) Training image. (b) One-dimensional representation obtained by using FIJI. (c)Training skeleton}\label{z}
		\end{figure}

SKESIM starts with a 2D binary training image, as presented in figure \ref{z}(a). One-dimensional representation is obtained by erosion (figure \ref{z}(b)), preserving the connectivity of the branches. The bifurcation points are found and the training skeleton results by connecting that points (nodes) by straight lines (edges), as can be seen in Figure \ref{z}(c).



\begin{figure}[h!]
	  \vspace{1cm}
		\centering
		\includegraphics[width=6.0cm]{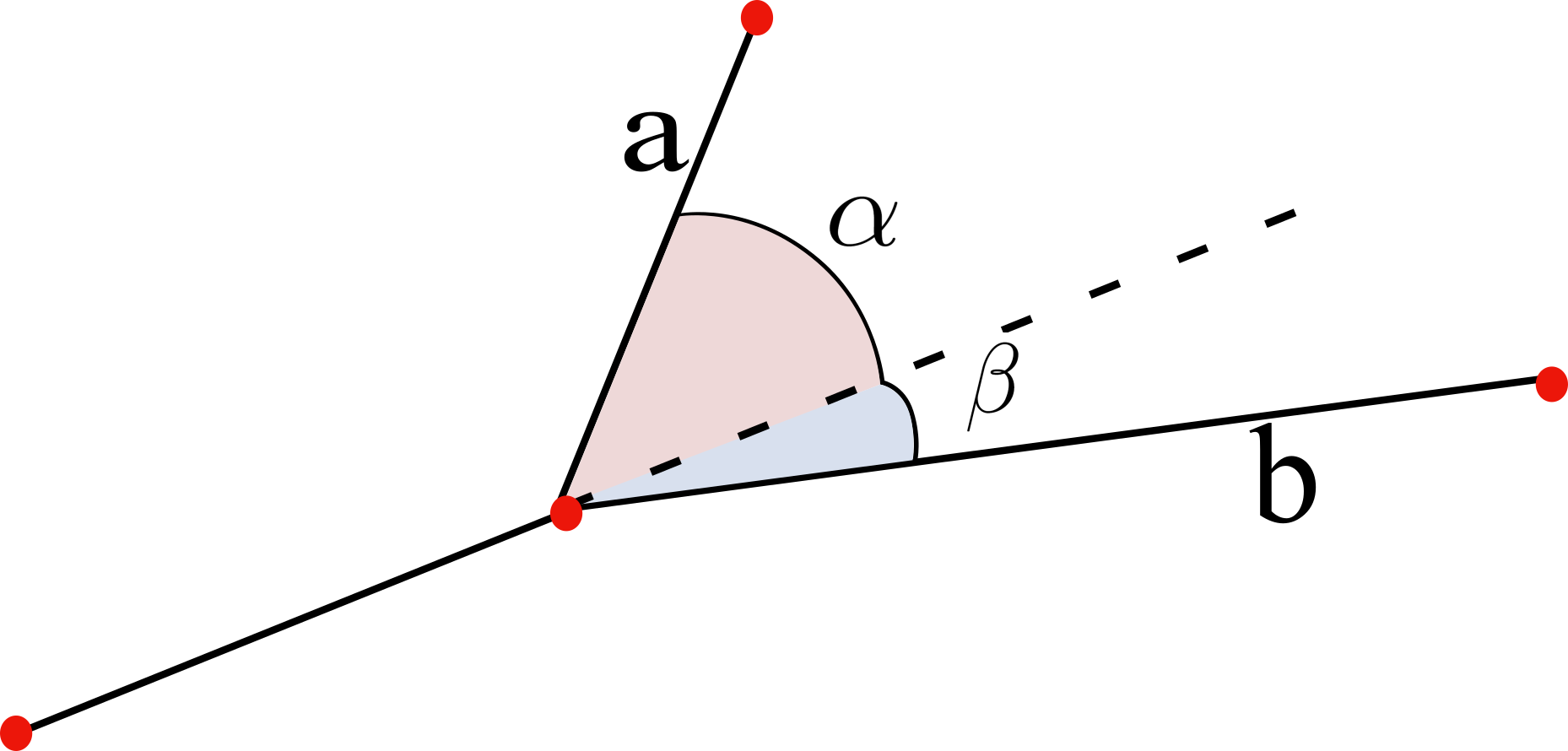}
		\caption{Bifurcation angles and lengths.}\label{bif_ang}
		\end{figure}

Two information are extracted from the skeleton to guide the generation of others skeletons: the bifurcation angles and the length of the edges. In figure \ref{bif_ang} can be seen the bifurcation angles $\alpha$ and $\beta$ of the edges $a$ and $b$, respectively. This information is used to obtain probability distribution functions for the angles and lengths, that will be used to simulate new skeletons.

\addcontentsline{toc}{subsection}{Skeleton definition}
 \subsection*{Skeleton definition}


A skeleton $Sk$ is given by two sets: the edges $E$ and nodes $N$. Each edge $e \in E$ is a directed segment defined by two points in $\mathbb{R}^2$, the father point $f_e$ and the son point $s_e$. The edge goes from the initial point $ f_e $ to the final point $ s_e $. 

A node $n \in N$ is an object formed by 3 elements: a point $p_n \in \mathbb{R}^2$, a vector $\alpha_n \in \mathbb{R}^2$ and an integer $M_n \in \mathbb{N}$. $\alpha_n$ contains the directions of the edges that arrive at the point $p_n$, it can be one or two edges. $M_n$ is the number of edges that arrive or left the point $p_n$, it is called the \textit{mark} of the node. We define $M_n$ as being a value in the set $\{1,2,3\}$.

\addcontentsline{toc}{subsection}{Skeleton synthesis}
 \subsection*{Skeleton synthesis}

 
The construction of the skeleton aims to mimic the channel system development. The idea is to generate a sequence of sucessive skeletons

\[Sk_0, Sk_1, Sk_2, \ldots, Sk_n, \ldots \]

where each skeleton $Sk_{m+1}$ is obtained by bifurcating the previous skeleton $Sk_m$, through the application of the function $Bif: SKE \rightarrow SKE$, where $SKE$ is the set of all the skeletons. We have that:
 
\[ Bif(Sk_m) = Bif^{m+1}(Sk_0) = Sk_{m+1}.\]

The function $Bif$ bifurcates nodes with mark 1 or 2. If the node has mark 1 then it is bifurcated in two channels, if it has mark 2, only one channel is generated.  The angle and the length of the new edge are chosen using the probability distributions previously extracted.

\begin{figure}[h!]
	  \vspace{1cm}
		\centering
		\includegraphics[width=8.5cm]{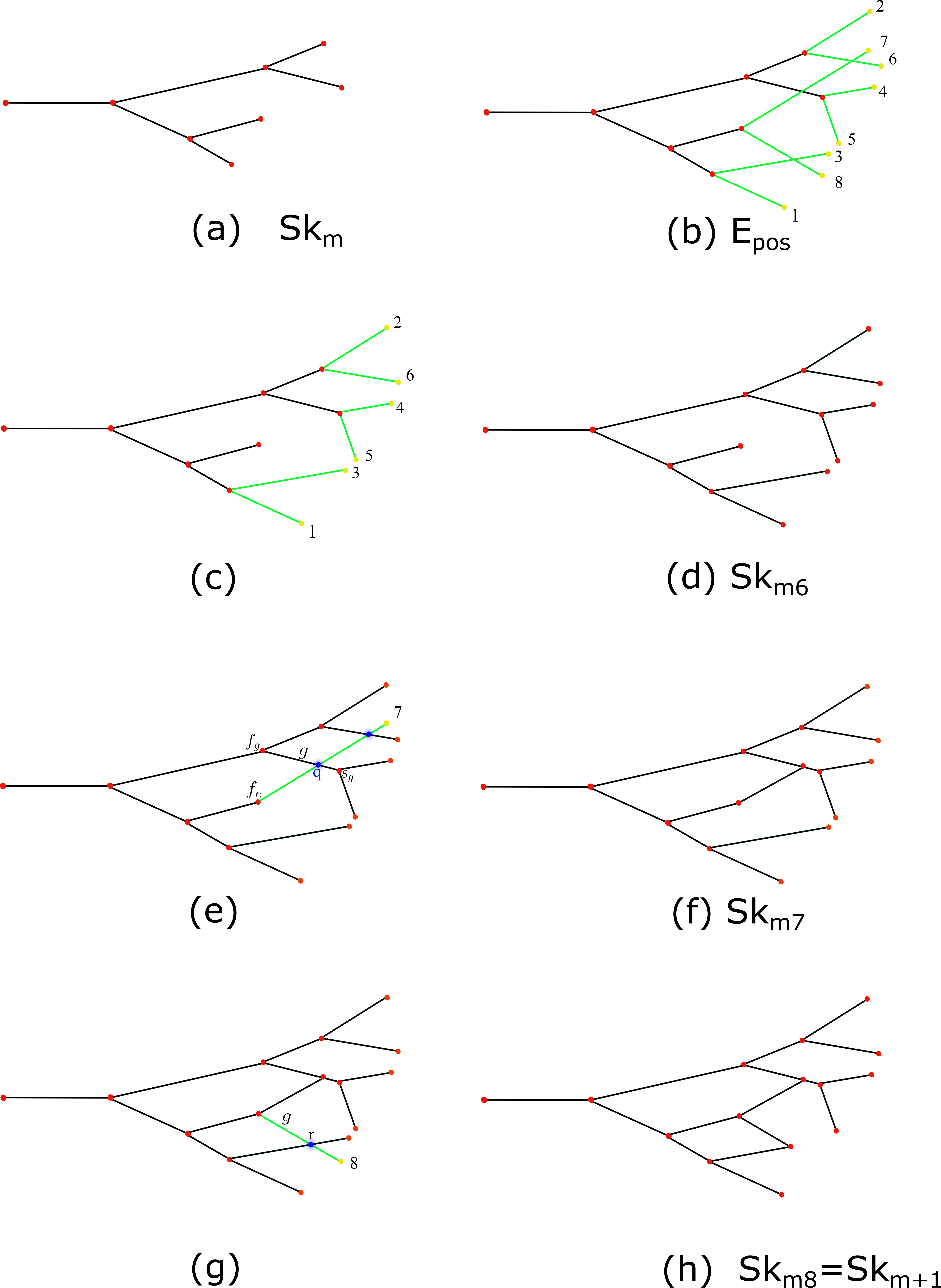}
		\vspace{0.5cm}
		\caption{Skeleton bifurcation.}\label{ske_syn}
		\end{figure}

Given a skeleton $Sk_{m}$, its set of nodes $N_m$ is traversed and the nodes with mark different to 3 are bifurcated generating new edges. All the resulting edges are stored in a set denoted by $E_{pos}$ of the possible edges to be included in the new skeleton $Sk_{m+1}$. In figure \ref{ske_syn}(b) the green edges form the set $E_{pos}$ from the skeleton in the figure \ref{ske_syn}(a). This set is traversed in a random order, that in the figure \ref{ske_syn} is given by the number at the end of each edge. Every time one edge in $E_{pos}$ is analyzed, another skeleton is created by adding the new edge.

The insertion process to generate $Sk_{m,n}$ from $Sk_{m,n-1}$ is not just adding the $n$-th edge, since this edge could intersect the already generated skeleton. If it has not a intersection then it is included as can be seen in the figure \ref{ske_syn}(c,d) where the edges have no intersection with any one. When the edges have intersections, the resulting skeleton is shown in figure \ref{ske_syn}(f,h).

The bifurcation process continues until some threshold is attained. This threshold could be the number of times the function $Bif$ is applied. Moreover, if the skeleton is generated within a specific region then its growth will be limited to the contour of this area.

\section{3D turbidite channels simulation}
 
\begin{figure}[h!]
\center
\includegraphics[width=7.5cm]{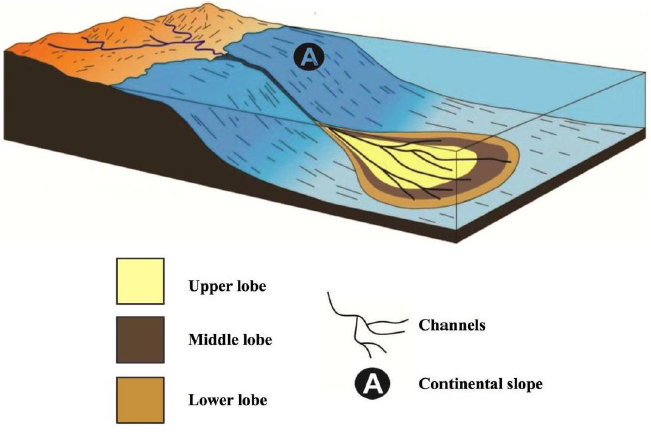}
\caption{Depositional system model of turbidites.}\label{lobe_example}
\end{figure}

The first step in the 3D modeling is to create the lobes, inside which the channels will be placed, as seen in figure \ref{lobe_example}. It is adopted the turbidity lobe modeling proposed in the work \cite{Alzate}. This is a simple depositional model with three turbidites lobes. Once the lobe has been created the skeleton synthesis is performed inside it. This means that the skeleton development is constrained by the lobe boundary.

\addcontentsline{toc}{subsection}{Lobe Modeling}
\subsection*{Lobe Modeling}

\begin{figure}[h!]
\center
\includegraphics[width=11.0cm]{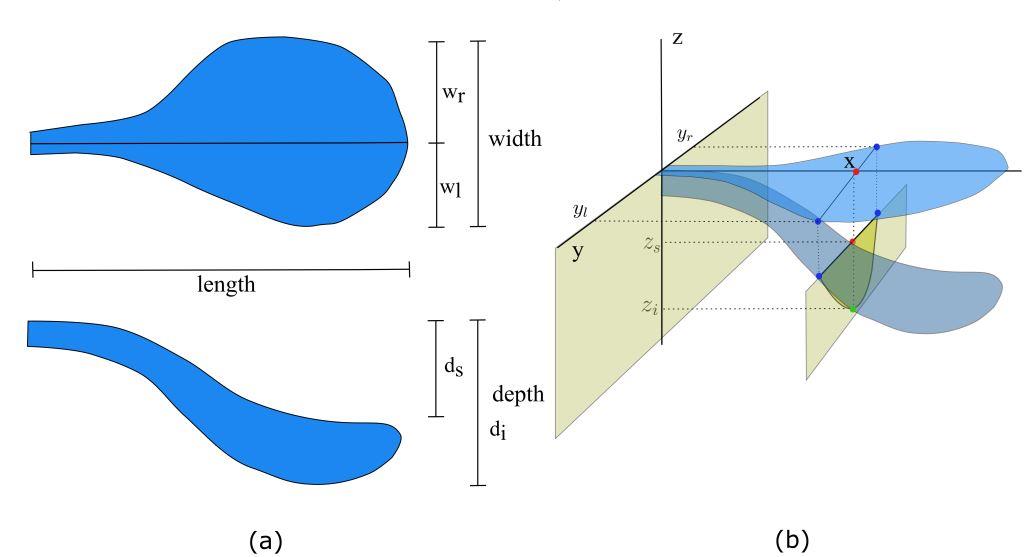}
\caption{Lobe parameters}
\label{parametros}
\end{figure}

The turbidite lobe is constructed using basically three parameters: depth, width and length. In the figure \ref{parametros}(a) the general structure given to the lobe and parameters used can be observed. 

\begin{figure}[h!]
\center
\includegraphics[width=8.0cm]{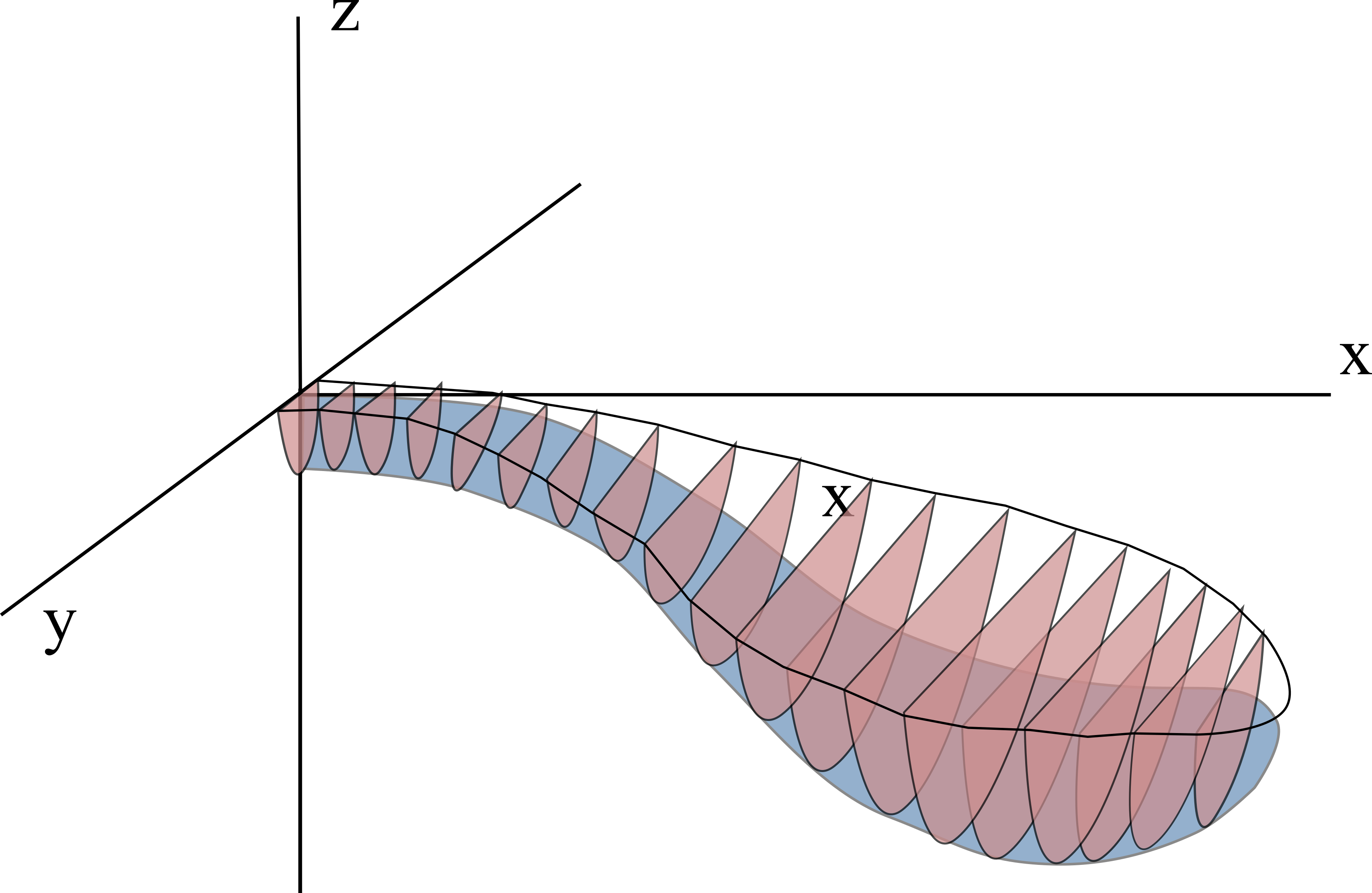}
\caption{Volume from the B-spline curves as the union of quarter of ellipses}
\label{vol}
\end{figure}

The algorithm creates the lobe by first creating two regions, one in the plane $xy$ and another in the plane $xz$ (see figure \ref{parametros}(b)), so these regions should coincide with the projections of the lobe to those planes. The regions are created using B-spline curves. These curve determine the volume of the lobe. Those curves are connected by quarters of ellipses and the volume is obtained joining the set of ellipses (see figure \ref{vol}).

The lobe is represented in a 3D grid. To determine whether a cube in the grid is inside a lobe or not, its center is evaluated in the equations that determine the lobe.

\addcontentsline{toc}{subsection}{Turbidite Channel Modeling}
\subsection*{Turbidite Channel Modeling}

\begin{figure}[h!]
\center
\includegraphics[width=8.5cm]{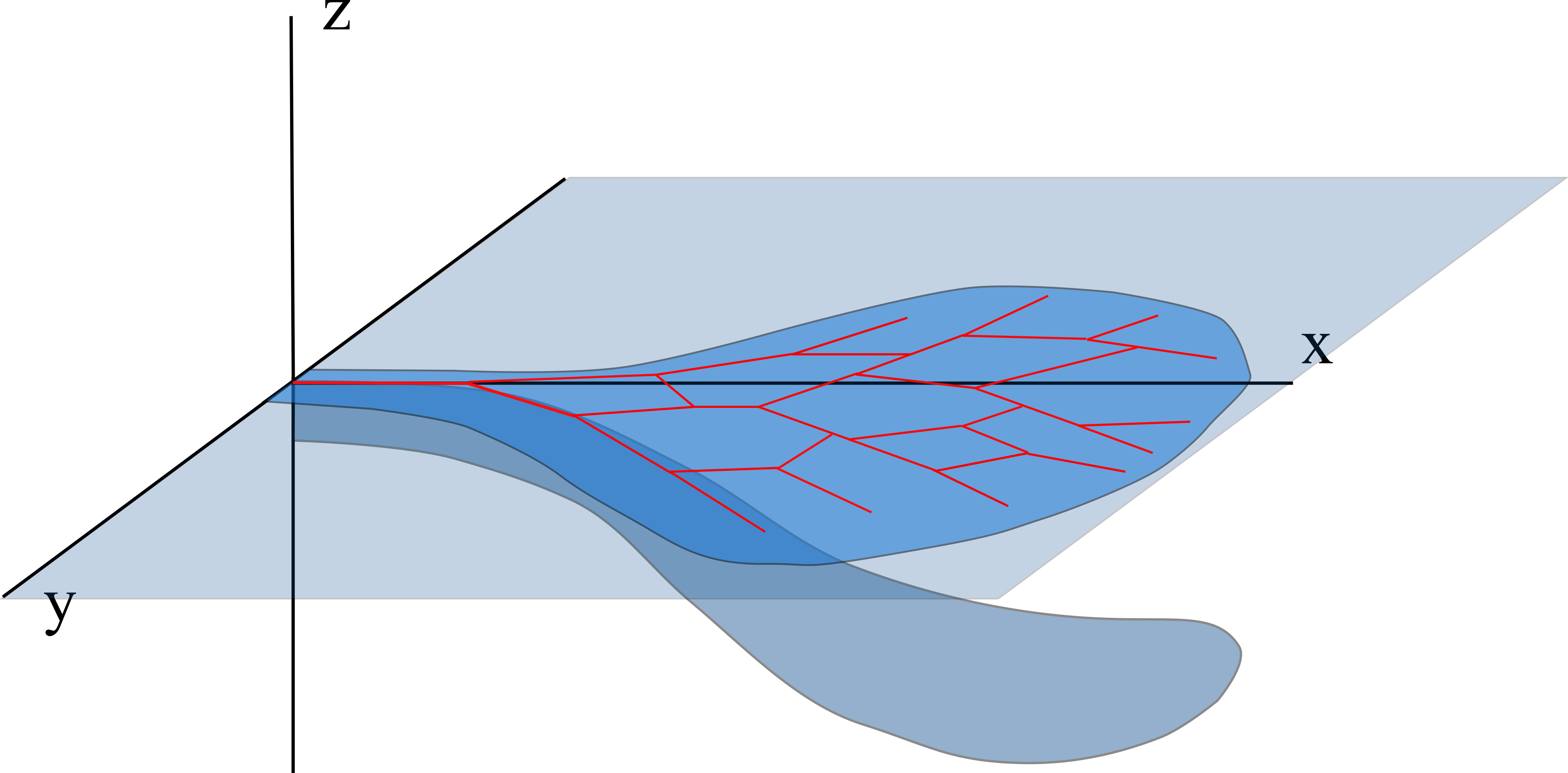}
\caption{Skeleton built inside the region limited by the curves $B_{right}$ and $B_{left}$ in the plane $xy$.}
\label{ske_lobe}
\end{figure}

Now, the skeleton is used to construct a 3D model of a turbidite channels. The basic idea is to built the skeleton in the plane $xy$ conditioned to the lobe limits \ref{ske_lobe}. The 3D channel system is built by giving volume to each edge in the edge set $E$ of the skeleton $S$. Each edge is used as the center line of a channel and the cross section orthogonal to the edge is a half-ellipse.

Until now the channels system has its surface in the $xy$ plane. To fit it inside the lobe, each point $(x,y,z)$ in it is translated downward, projecting the former channels system to the superior surface of the lobe.

\section{Simulations}

\begin{figure}[h!]
		\centering
		\includegraphics[width=8.5cm]{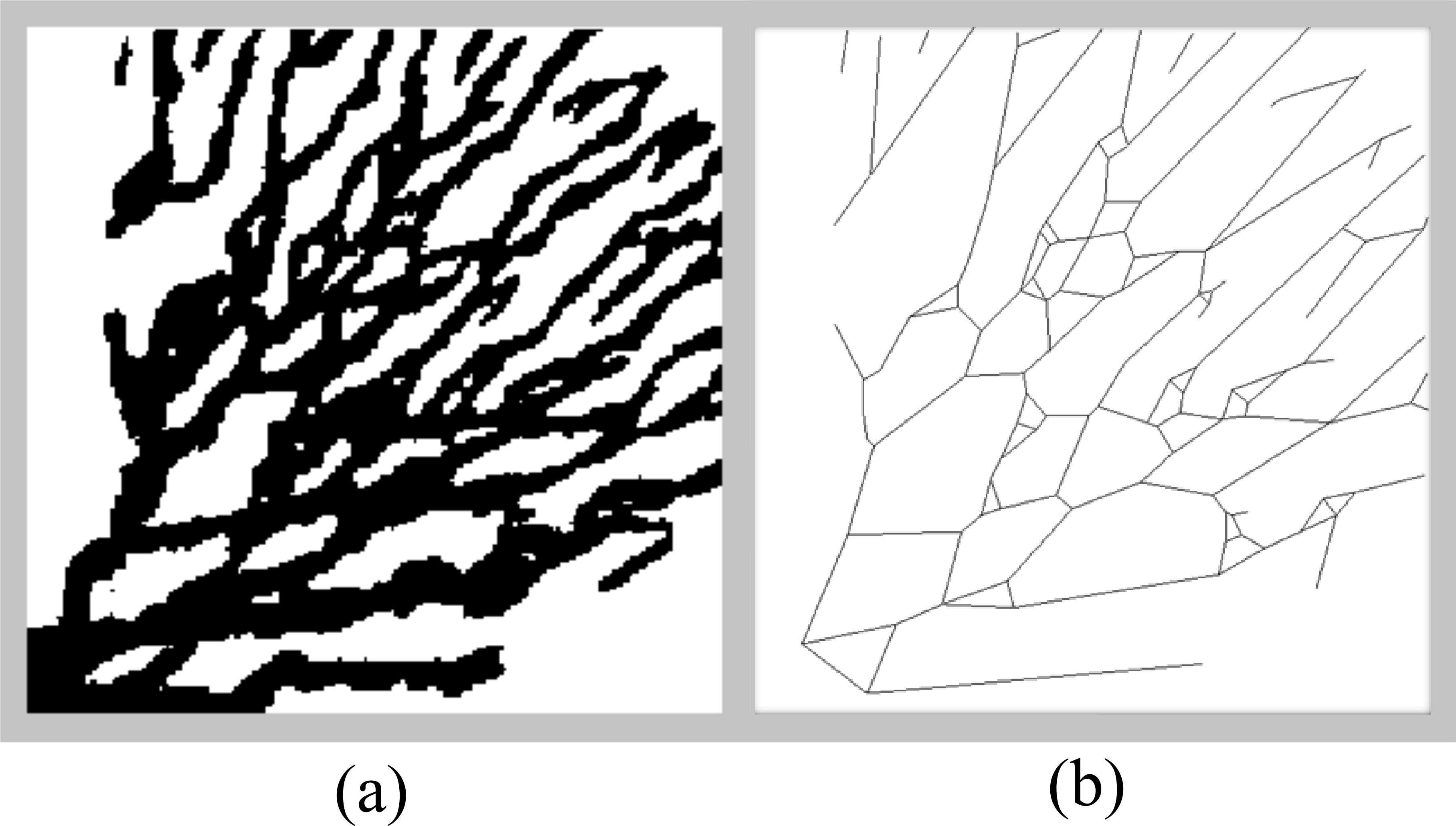}
		\caption{(a) Training image. (b) Training skeleton.}\label{x}
		\end{figure}

The training image used is given by figure \ref{x}(a). From this image, the training skeleton is obtained and it is illustrated in figure \ref{x}(b). This training skeleton is the source of information used to generate others skeletons. 

\begin{figure}[h!]
	 	\centering
		\includegraphics[width=9.0cm]{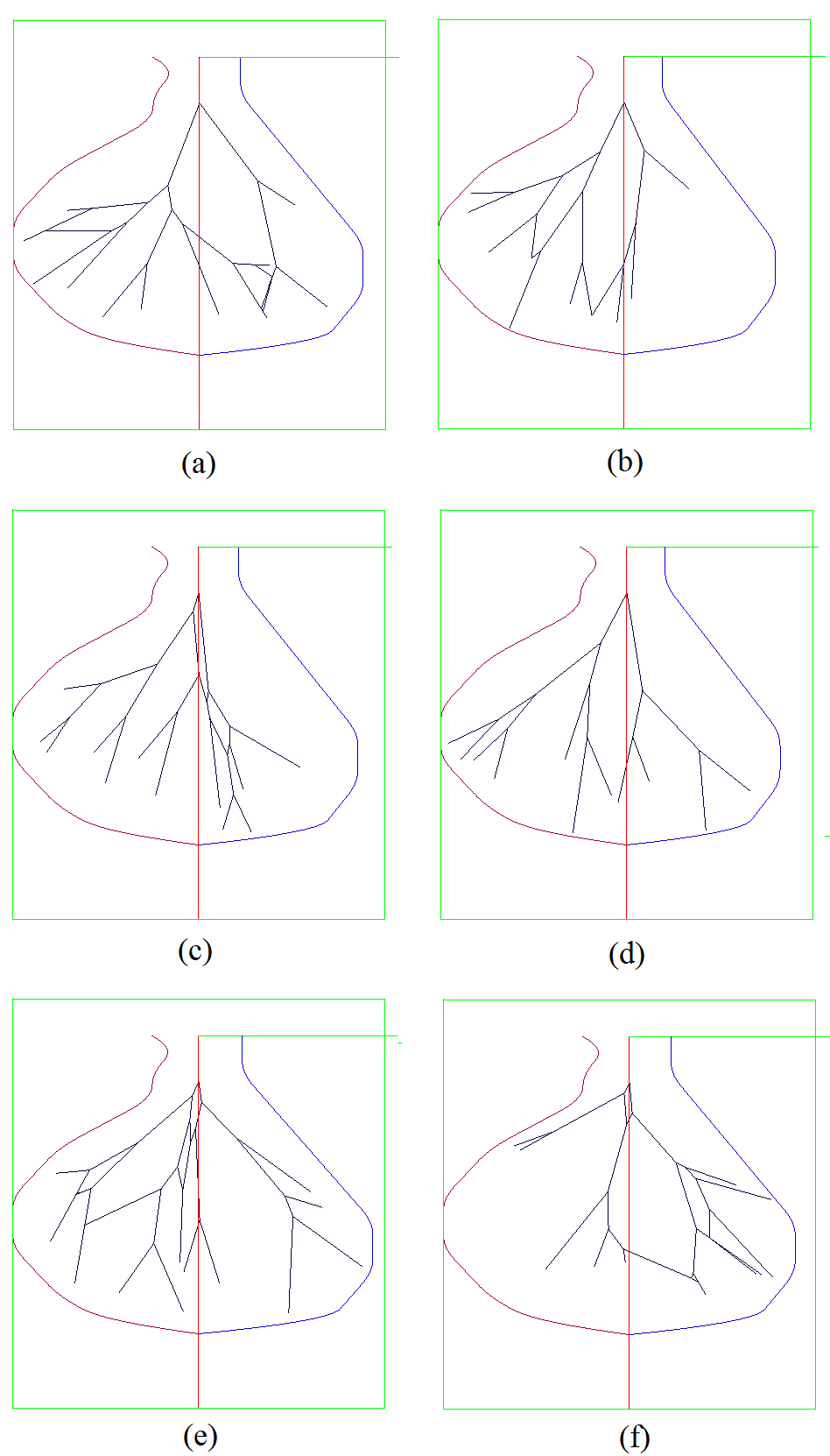}
		\caption{Six skeletons obtained using the training skeleton \ref{x}(b)}
		\label{realiz_ske_ex1}
		\end{figure}

From the training skeleton given by the figure \ref{x}(b), a uniform distribution of the bifurcation angles and channel lengths are obtained and used to generate others skeletons. These new skeletons are constructed following the algorithm explained in the previous section. In the figure \ref{realiz_ske_ex1} six simulations of skeletons are presented. The B-spline curves in plane $xy$ that limit the skeleton can be appreciated too.

\newpage

\begin{figure}[H]
\center
\includegraphics[width=10.0cm]{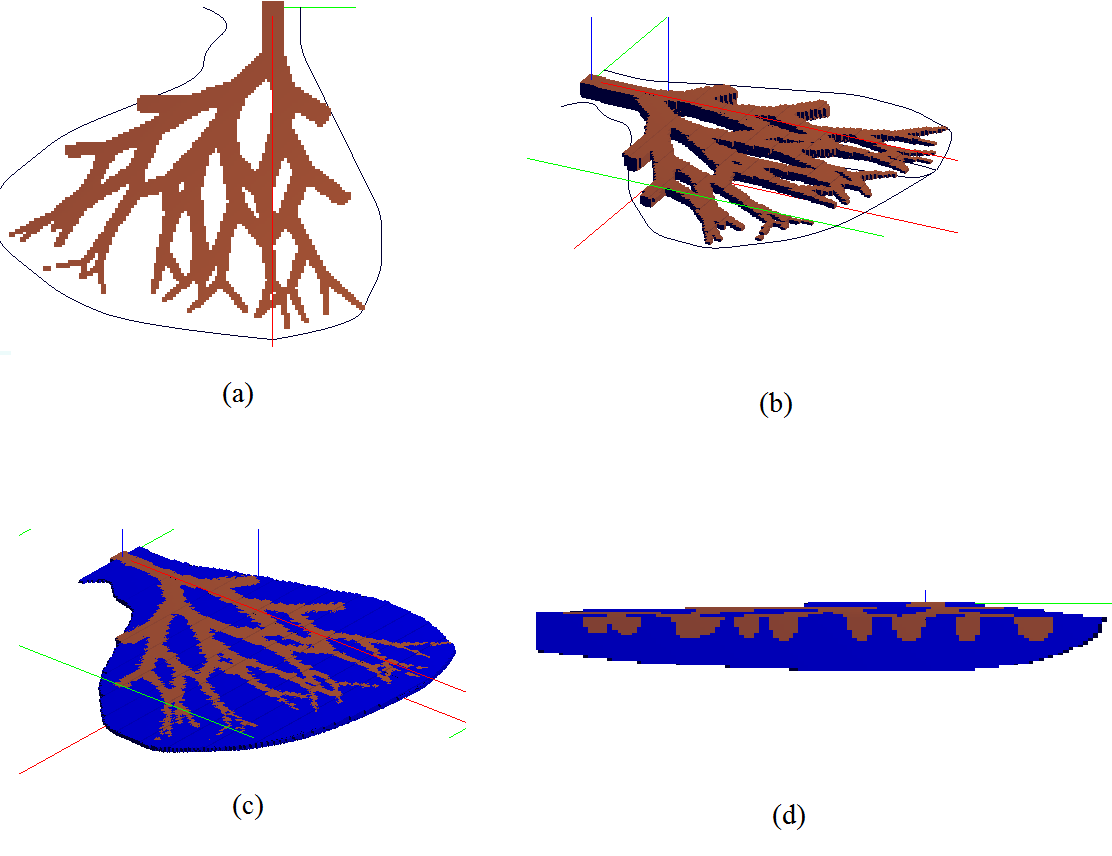}
\caption{One channel system simulation.}\label{Realiz_1}
\end{figure}

From the skeletons, that lie in the $xy$ plane, a 3D model is generated. In the figure \ref{Realiz_1} a simulation of one lobe and one channel system is presented. Figures \ref{Realiz_1}(a, b) show only the channel system, in (a) an upper view is shown. In the figure \ref{Realiz_1}(c), the same channel system is illustrated inside the lobe. The figure \ref{Realiz_1}(d) shows a cross-section, perpendicular to the $x$ axis, of the total system, here the ellipsoidal form of the channels is appreciated. 

\begin{figure}[H]
\center
\includegraphics[width=10.0cm]{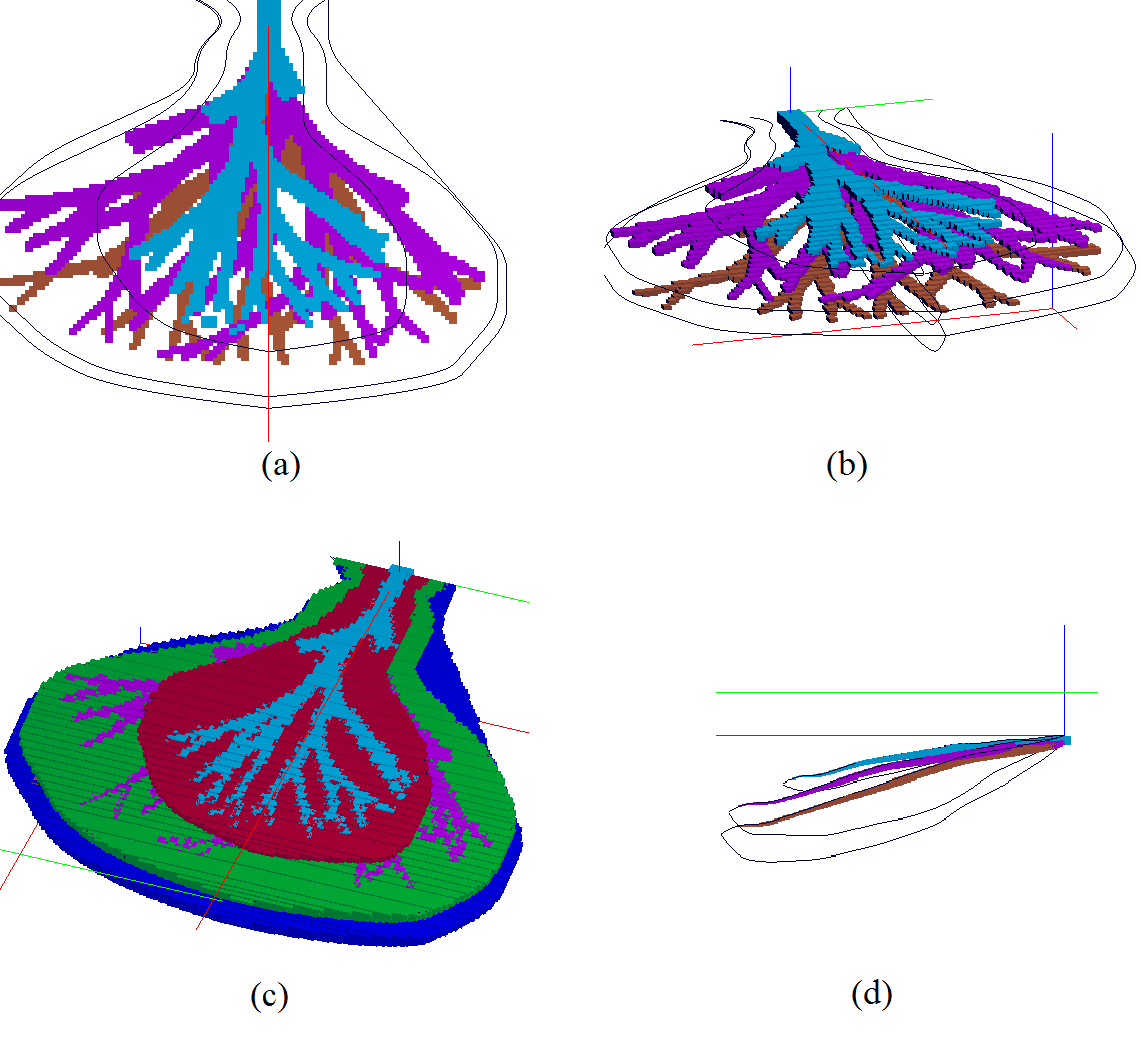}
\caption{Three channels system simulation.}\label{Realiz_5}
\end{figure}

In the figure \ref{Realiz_5}, three channel systems were simulated inside three lobes. In the figure \ref{Realiz_5}(a) an upper view of the channel set is presented. A different perspective of the same system is shown in the figure \ref{Realiz_5}(b). Together with the channels, the lobes that contains them are drawn in the figure \ref{Realiz_5}(c). In the figure \ref{Realiz_5}(d), a lateral view of the three channels is shown. In this same image it can also be observed the b-spline curves in the $xz$ plane (these curves define the depth of the lobe). 

\begin{figure}[H]
\center
\includegraphics[width=10.0cm]{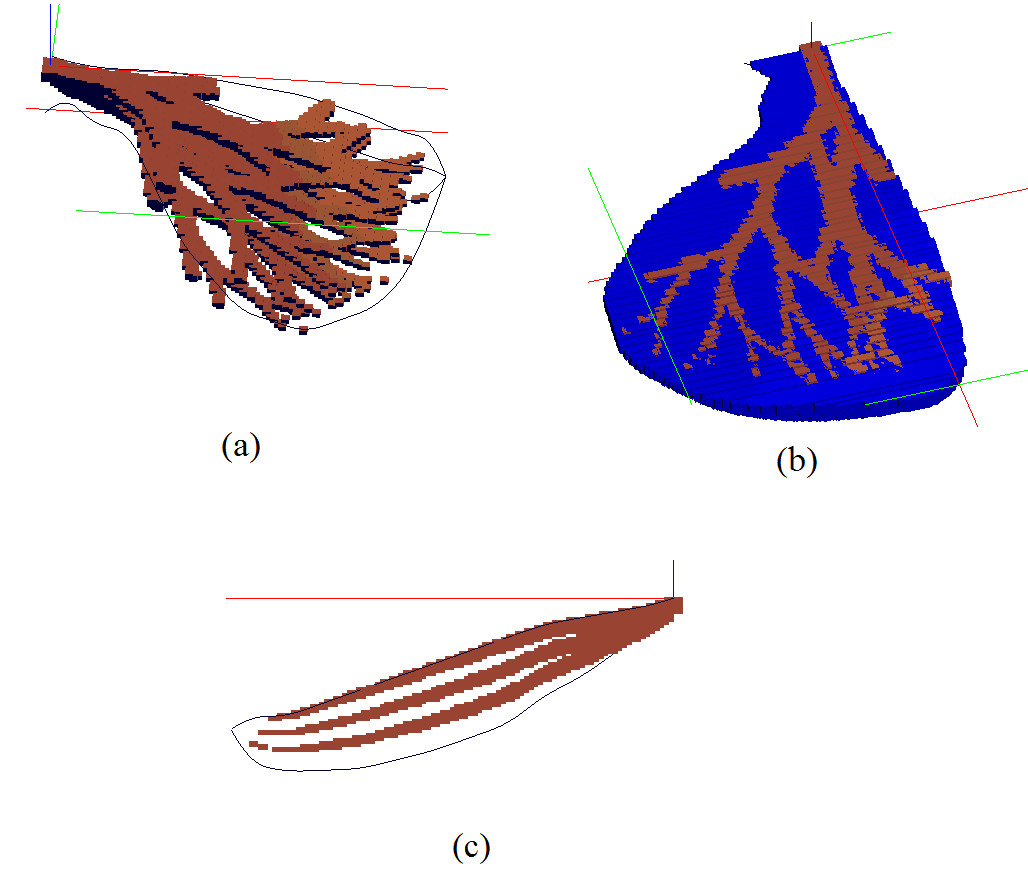}
\caption{Three channels system simulation in one lobe.}\label{Realiz_13}
\end{figure}

In the previous realizations, in each lobe one channel system was generated. The SKE-SIM simulation can be easily modified to allow the generation of more than one channel system in the same lobe. In the figure \ref{Realiz_13}, three channels systems were simulated inside one lobe. In figure \ref{Realiz_13}(a), only the three channels are visualized. In figure \ref{Realiz_13}(b) the system is shown together with the lobe that contains them, in this case only the system in the surface of the lobe can be seen. In the figure \ref{Realiz_13}(c), the lateral view  (parallel to the plane $xz$) shows how the three systems are distributed inside a lobe, the black curve surrounding the systems are the b-spline curves in the plane $xz$. 


 \section{Conclusions}

A 3D object-based modeling for turbiditic channels inside a lobe was presented. The lobe is created using the work of \cite{AY}. To build the channel system we use a tree-like image that represents the projection to the plane $xy$ of the channel system. This is the training image and contains the basic geometry desired in the object. A linear approximation of this image is obtained and used to extract two probability distributions for the parameters values (bifurcation angles and length), which are employed to generate new skeletons.

The generation of the skeletons is very simple. The skeleton is constructed sequentially. In each step, a couple of new edges emerge from each of the final nodes. The directions and length of those edges are determined using the probability distributions. Some rules are applied for especial edges or when an intersection occurs. The skeleton continues growing until a threshold is reached. Here, the boundary of the projection of the lobe to the plane $xy$ was used as constraint.

Once the skeleton is simulated, the 3D channel system is created inside the lobe. The idea is to give a volume to the skeleton, each edge is thickened, deepened and translated downward such that it lies inside the lobe. SKE-SIM is not a complex method, it does not try to mimic the detailed physical behavior of the phenomena. Nevertheless, the results are visually appealing. 

There are two fundamental assumptions upon which the method is supported: the geometry of the channel system can be reasonably represented by a thickening of a unidimensional structure, the skeleton. Secondly, the distance traversed by a channel before something happens making it bifurcates is controlled by a random variable whose probability distribution does not depend on the position of the channel. This also happens for the bifurcation angles. Thus, these two quantities can be simulated using the same random variables (one for each quantity) no matter where the phenomenon occurs.


\bibliographystyle{unsrt}
\bibliography{Paper_v8}

\end{document}